\documentclass[letterpaper, 10 pt, conference]{ieeeconf}  
\IEEEoverridecommandlockouts 
\usepackage{graphics} % for pdf, bitmapped graphics files
\usepackage{epsfig} % for postscript graphics files
\usepackage{mathptmx} % assumes new font selection scheme installed
\usepackage{times} % assumes new font selection scheme installed
\usepackage{amsmath} % assumes amsmath package installed
\usepackage{amssymb}  % assumes amsmath package installed
\usepackage{booktabs}    % \toprule, \midrule, \bottomrule
\usepackage{multirow}    
\usepackage{array}      
\usepackage{makecell}
\title{\LARGE \bf
Miniaturized Pneumatic Actuator Array\\ for Multipoint Deep Pressure Tactile Stimulation
}

\author{
    Ava Chen, 
    Megan C. Coram,
    Cosima du Pasquier, and
    Allison M. Okamura%
    \thanks{This work was supported in part by the National Institutes of Health: National Institute of Biomedical Imaging and Bioengineering under grant R21EB036190. The authors are with the Department of Mechanical Engineering, Stanford University, Stanford, CA 94305, USA. {\tt\small \{avachen, mccoram, cosimad, aokamura\}@stanford.edu}}%
}

\begin{document}

\maketitle
\thispagestyle{empty}
\pagestyle{empty}

%%%%%%%%%%%%%%%%%%%%%%%%%%%%%%%%%%%%%%%%%%%%%%%%%%%%%%%%%%%%%%%%%%%%%%%%%%%%%%%%
\begin{abstract}
Wearable distributed tactile devices aim to provide multipoint touch stimuli, but struggle to provide sufficient forces (\textgreater~1~N) at frequencies to invoke deep pressure sensation with minimal encumbrance at small scales. This work presents a method of fabricating arrays of pneumatic actuators from thermoplastic-coated textiles. By routing pneumatic inlets to a common fold line in the fabric, we demonstrate that multiple pneumatic pouch actuators can be formed in a single simple heat-pressing operation that does not require the use of sacrificial blocking layers. The method accommodates a range of actuator diameters and spacing distances, including as compact as 8~mm diameter actuators spaced 1~mm apart, which enables use in fingertip wearable devices. In a blocked force test, these small pneumatic textile actuators exert 2.1 N when pressurized to 230~kPa. With this pair of actuators, we demonstrate an example application in which we invoke both distinct and summative stimuli, suggesting the possibility of titrating just noticeable difference in amplitude with a textile actuator array.
\end{abstract}

\section{INTRODUCTION}
Textile-based pneumatic pouch actuators are attractive for wearables because they feature thin, lightweight, and flexible form factors that provide force stimuli to the skin at a range of frequencies (including static stimuli), and can route their power sources away from the point of stimulation. The ability of pneumatic systems to statically deform the skin offers a compelling alternative to vibration-based systems when designing haptic feedback devices that prioritize spatial acuity. We seek to use multipoint pneumatic actuators as part of developing a wearable sensory prosthesis for individuals with PIEZO2 Loss of Function (PIEZO2-LOF), a congenital condition in which proprioception and vibration senses are absent but deep tissue sensation (\textgreater 1 N) is preserved~\cite{case2021}. Our goal is to provide proprioceptive feedback cues via sensory substitution at the fingertips, wrist, and forearm. Although textile pneumatic actuators have been previously explored in a range of sizes and geometries~\cite{jumet2023, coram2025preprint}, the combination of tactor miniaturization, dense spacing, and sufficiently high force output has not yet been achieved to realize multipoint deep pressure stimulation at a small scale. 

This work presents a heat-press method of manufacturing single-layer, multi-actuator arrays from TPU-coated nylon. Here we focus on a fingertip-sized array; we measure force generation  and investigate the feasibility of creating distinct and summed stimuli.

\section{Pneumatic Actuator Array}
\begin{figure}[t]
    \centering
    \vspace{2mm}
    \includegraphics[width=0.98\linewidth]{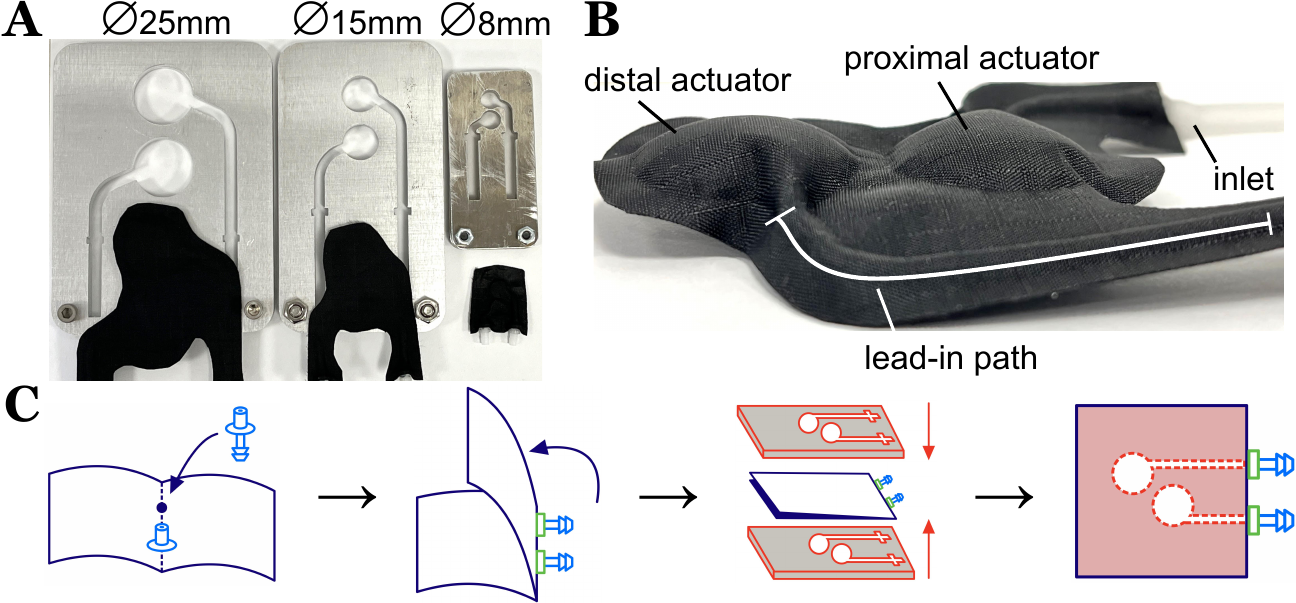}
    \vspace{-3mm}
    \caption{A: The textile pneumatic actuators and molds produced in this paper. B: Annotated photo of actuator array. C: Fabrication process.}
    \label{fig:fabrication}
    \vspace{-5mm}
\end{figure}

The working principle is to heat-seal fabrics impregnated with a thermoplastic film in order to form single-layer inflatable bladders. Here, we focus on demonstrating multipoint capabilities by producing 2$\times$1 fabric actuator arrays, shown in Figs.~\ref{fig:fabrication}A--B, that can be either assembled into garments or installed as patches in preexisting wearables.

\begin{figure*}[t]
    \centering
    \includegraphics[width=\textwidth]{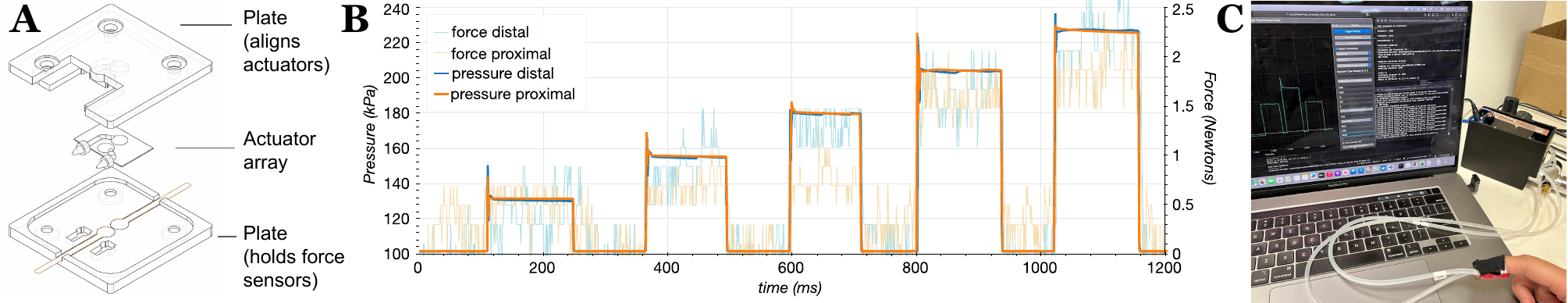}
    \caption{A: Force testing setup. B: Measured force and pressures for distal and proximal actuators (relative to air inlets). C: Perceptual task setup.}
    \label{fig:blockedforce}
    \vspace{-4mm}
\end{figure*}

Our pneumatic actuators are made from 70~denier ripstop nylon with a single-sided TPU coating (Quest Outfitters~\#1056). We fold a sheet of this fabric in half with the TPU-coated side facing inward, then use sharp tweezers to pierce holes at the fold line for air inlets (Fig.~\ref{fig:fabrication}C). We insert 2~mm~ID polypropylene barb fittings (Gikfun \#EK1260) into these holes, and clamp the ragged edge of the punctured fabric against the nozzle using a thin collar of silicone tubing in order to form a seal. With fittings inserted, we place the folded fabric part into a custom aluminum mold, which consists of two identical 3~mm thickness aluminum plates. Each plate is lasercut with voids for the desired actuator array geometry, including cutout features to use the barb fittings for alignment. When heat pressed (DabPress Commercial Rosin Press), the high heat conductivity of aluminum relative to air seals only the regions of the TPU-coated fabric that are in contact with the molds, without the need for blocking material. Following prior characterization of pressure and temperature effects on bond strength for this fabric~\cite{coram2025preprint}, we set heat press parameters to 127~$^\circ$C temperature, 5 MPa pressure, 90 second duration for all samples. The resulting actuators can sustain a maximum inflation pressure of 375~kPa before failure (about 275~kPa above atmospheric pressure).

In this work, we produced three prototypes that demonstrate a range of possible actuator diameters and spacings. We focused experiments on investigating the capabilities of a fingertip-sized array, for which we selected an 8~mm actuator diameter and 1~mm edge-to-edge spacing, with resulting indentation depth of 2.5~mm. For demonstration purposes, we also produced textile arrays with 15~mm and 25~mm diameters, both with 5~mm spacings, to match previous non-textile multipoint tactors in the literature~\cite{Kodali2024}. We chose circular tactor geometries to minimize stress concentrations at the edges and maximize indentation depth, and used long, thin lead-in paths to position inlet hose connections away from the main contact areas of the actuators. Our fabrication method required all air hose connections to be placed co-linearly, although optional post-process cutting steps could enable three-dimensional inlet routings.

\section{Preliminary Results}
\subsection{Blocked Force Characterization}
We assessed the performance of the fingertip-sized actuator using a blocked force test adapted from Coram et al.~\cite{coram2025preprint}.
As shown in Fig.~\ref{fig:blockedforce}A, we sandwiched the actuator array between 3D-printed plates that constrain expansion without impeding airflow through the inlet connections. When inflated, each actuator in the array pressed against a force sensor (Singletact S8-100N). The actuator array was inflated by a pneumatic controller to a series of absolute pressures from 130 to 230 kPa at 25 kPa intervals (5 total peaks), held for approximately 100 ms, then vacuumed to atmospheric pressure between each pressure setpoint. Fig.~\ref{fig:blockedforce}B shows the force-pressure correspondence from a single sample.

The pneumatic controller used in this study provided a maximum pressure of 230 kPa, which yielded over 2.1~N of force transmission. This blocked-force plate setup over-represents actuator performance by directing all deformation along the axis of the force sensor; realistic conditions when embedded in a soft wearable would yield smaller forces.

\subsection{Just Noticeable Difference: One vs. Two Contacts}

We compared the just noticeable difference (JND) in amplitude given one or two pressure stimuli from the fingertip actuator array, following a three-alternative forced choice \mbox{(3-AFC)} method adapted from Pat\'e et al.~\cite{pate2024}. We performed preliminary tests with two participants: one co-author and one external user. The actuator was embedded in an elastic finger cap and was placed at the pad of the user's dominant index finger. The hand and elbow were supported while the finger was held in the air. We did not take steps to block potential auditory or visual feedback from the actuators, although the volume of air was small and the actuators were hidden in the garment. At the start of both tests, we acclimated the user by inflating the device (distal actuator for one-contact) and verified that all tactors were perceived as recognizably distinct contact points. Fig.~\ref{fig:blockedforce}C demonstrates the setup, including the laptop's control display that is hidden from the user during the test.

We presented a row of three sequential square wave signals, which inflated the actuator to a given pressure amplitude for a 2~second duration then depressurized for a 0.5~second pause. Two signals were identical and one ``alternative'' was different. The alternative signal was initially set to 5~kPa higher amplitude than the 120~kPa reference. Users were asked to verbally identify which stimulus was different in each set of three (the order of the alternative stimulus was randomized). Step sizes proceeded according to a modified rapid staircase algorithm~\cite{pate2024}. Upon convergence, we estimated JND by taking the average amplitude differences for the last right/wrong answer pair. For one user, the JND (9.6~kPa) was identical between one- and two-contact stimuli. For the other user, the one-contact stimulus had a higher JND (6.5~kPa) than the two-contact stimulus (4.1~kPa), which indicated potential presence of spatial summation at the fingertip. 

\section{Conclusions and Future Work}
We demonstrate that fabric actuators can be sufficiently miniaturized and densely placed for use in perceptual experiments at the fingertip. Additionally, these actuator arrays generate sufficient force to invoke deep pressure sensation. In future work, we will develop wearable devices using these actuators to provide substitutive haptic feedback from multiple points of stimuli, including at the forearm, wrist, and fingers. The ultimate goal of this work is to use these devices with individuals with PIEZO2-LOF to assist proprioception.

%%%%%%%%%%%%%%%%%%%%%%%%%%%%%%%%%%%%%%%%%%%%%%%%%%%%%%%%%%%%%%%%%%%%%%%%%%%%%%%%

\bibliographystyle{IEEEtran} % We choose the "plain" reference style
\bibliography{references} % Entries are in the refs.bib file

\end{document}